\documentclass[10pt, final, conference, twoside]{IEEEtran}
\usepackage{cite}
\usepackage{amsmath,amssymb,amsfonts}
\usepackage{array,multirow,graphicx}
\usepackage{tikz}
\usepackage{scalerel}
\usepackage{mathtools,eqparbox}
\usepackage{xcolor}
\usepackage{commath}
\usepackage{pgfplots}
\usepackage{subfigure}
\usepackage[hidelinks]{hyperref}
\usepackage[utf8]{inputenc}
\usepackage[T1]{fontenc}
\usepackage{psfrag}
\usepackage{flushend}
\def\Var{{\textrm{Var}}}
\def\E{{\textrm{E}}}

\newcommand{\muris}{\mu_{_{\mathrm{RIS}}}}
\newcommand{\C}{\cal{C}}
\newcommand{\rg}{R_{g}}

\newcommand{\ru}{R_{u}}

\newcommand{\lambdahaps}{\lambda_\mathrm{{HAP}}}
\newcommand{\lambdab}{\lambda_\mathrm{{B}}}

\newcommand{\HMIN}{H_{\mathrm{min}}}
\newcommand{\HRIS}{H_{\mathrm{RIS}}}
\newcommand{\HHAP}{H_{\mathrm{HAP}}}

\def\Var{{\textrm{Var}}}
\def\E{{\mathbb{E}}}

\newtheorem{theorem}{Theorem}

\newtheorem{lemma}{Lemma}

\newtheorem{proposition}{Proposition}
\newtheorem{optimization problem}{Optimization Problem}
\pgfplotsset{compat=1.14}

\tikzstyle{arrow} = [thick,->,>=stealth]
\tikzstyle{block} = [rectangle, rounded corners, minimum width=1cm, minimum height=1cm,text centered, draw=black, fill=red!30]
\tikzstyle{input} = [circle, minimum width=2.5cm, minimum height=1cm, text centered, draw=black, fill=blue!30]

\usetikzlibrary{svg.path}
\definecolor{orcidlogocol}{HTML}{A6CE39}
\tikzset{
    orcidlogo/.pic={
        \fill[orcidlogocol] svg{M256,128c0,70.7-57.3,128-128,128C57.3,256,0,198.7,0,128C0,57.3,57.3,0,128,0C198.7,0,256,57.3,256,128z};
        \fill[white] svg{M86.3,186.2H70.9V79.1h15.4v48.4V186.2z}
        svg{M108.9,79.1h41.6c39.6,0,57,28.3,57,53.6c0,27.5-21.5,53.6-56.8,53.6h-41.8V79.1z M124.3,172.4h24.5c34.9,0,42.9-26.5,42.9-39.7c0-21.5-13.7-39.7-43.7-39.7h-23.7V172.4z}
        svg{M88.7,56.8c0,5.5-4.5,10.1-10.1,10.1c-5.6,0-10.1-4.6-10.1-10.1c0-5.6,4.5-10.1,10.1-10.1C84.2,46.7,88.7,51.3,88.7,56.8z};
    }
}

\newcommand\orcidicon[1]{\href{https://orcid.org/#1}{\mbox{\scalerel*{
                \begin{tikzpicture}[yscale=-1,transform shape]
                \pic{orcidlogo};
                \end{tikzpicture}
            }{|}}}}
\newcounter{MYtempeqncnt}

\begin{document}

\title{Enhancing HAP Networks with\\Reconfigurable Intelligent Surfaces\\
}

\author{Islam~M.~Tanash\IEEEauthorrefmark{1}${\textsuperscript{\orcidicon{0000-0002-9824-6951}}}$, Ayush Kumar Dwivedi\IEEEauthorrefmark{2}${\textsuperscript{\orcidicon{0000-0003-2395-6526}}}$, Fatemeh Rafiei Maleki\IEEEauthorrefmark{2}, and Taneli Riihonen\IEEEauthorrefmark{2}${\textsuperscript{\orcidicon{0000-0001-5416-5263}}}$~\IEEEmembership{Senior Member}\\
\IEEEauthorrefmark{1}Department of Information and Communications Engineering, Aalto University, Finland\\
\IEEEauthorrefmark{2}Faculty of Information Technology and Communication Sciences, Tampere University, Finland\\
\texttt{islam.tanash@aalto.fi}, \texttt{ayush.dwivedi@tuni.fi}
}
\maketitle

\begin{abstract}
This paper presents and analyzes a reconfigurable intelligent surface (RIS)-based high-altitude platform (HAP) network. Stochastic geometry is used to model the arbitrary locations of the HAPs and RISs as a homogenous Poisson point process. Considering that the links between the HAPs, RISs, and users are $\kappa$--$\mu$ faded, the coverage and ergodic capacity of the proposed system are expressed. The analytically derived performance measures are verified through Monte Carlo simulations. Significant improvements in system performance and the impact of system parameters are demonstrated in the results. Thus, the proposed system concept can improve connectivity and data offloading in smart cities and dense urban environments.
\end{abstract}

\begin{IEEEkeywords}
High-altitude platforms (HAPs), reconfigurable intelligent surfaces (RISs), stochastic geometry.
\end{IEEEkeywords}

\section{Introduction}
The growing demand for wireless services, driven by the proliferation of mobile broadband and the Internet-of-Things (IoT), requires innovative performance solutions. Recently, non-terrestrial networks (NTN) have become popular as force multipliers with terrestrial networks \cite{3GPP-TR38.811:Study_NTN}. For example, satellites can provide wide-area coverage by establishing line-of-sight (LoS) communication due to their higher elevation angle and large beam coverage on ground. Since satellites come with high deployment and maintenance costs, high-altitude platforms (HAPs) have emerged as a promising technology for better agility and cost-effectiveness \cite{Alam-MCOM2021:High,Alfattani-MVT23:Multimode, ITU-HAPS}. Operating at altitudes from 20 to 50 km \cite{3GPP-TR38.821:Solutions_NTN}, HAPs can complement terrestrial networks by offering enhanced connectivity, especially in underserved or remote areas. Their mobility and form factor allow for rapid deployment, making them suitable for dynamic and temporary network enhancements. However, they face challenges, particularly in dense urban environments where buildings and other structures obstruct LoS communication. 

Reconfigurable intelligent surfaces (RISs) provide a solution to this challenge \cite{Tekbıyık-MVT22:Reconfigurable}. A RIS can manipulate electromagnetic waves to improve signal propagation, creating virtual LoS paths where direct communication is obstructed. By integrating RIS with HAP, we can enhance signal coverage and reliability, particularly in environments with high user density or significant structural obstacles. As described in \cite{Ye-JPROC22:Nonterrestrial}, RIS can be integrated with HAP networks in two ways. The first approach---aerial RIS (ARIS)---involves having the RIS onboard a HAP as a payload. The second approach---terrestrial RIS (TRIS)---places the RIS on terrestrial objects, e.g., rooftops or sides of buildings, facilitating the connection between the HAP and the user via the RIS. Each method has advantages and limitations, and the selection can usually be made based on the application in mind. However, having an RIS onboard a HAP can make the platform bulky, requiring higher payload capacity concerning aerodynamics. It also leads to increased energy consumption and reduced airtime. Conversely, placing the RIS on existing terrestrial infrastructure can keep the HAP more agile and extend its flying time. Terrestrial-based RIS can serve as a multipurpose infrastructure, supporting HAP-based and other terrestrial wireless communications, thereby increasing the system's reusability.

Recently, there have been several efforts towards standardizing HAP networks at the 3rd Generation Partnership Project (3GPP). Initially, a study was conducted to identify potential enhancements for the aerial user equipment, specifying several improvements in Release 15 \cite{3GPP-TR36.777:Study_UAV}. Subsequently, solutions for 5G New Radio (NR) supporting airborne platforms were presented in \cite{3GPP-TR38.821:Solutions_NTN}, which were later transformed into normative requirements for non-terrestrial networks (NTN) in Release~17. On the other hand, RIS still needs to be a study item in 3GPP and some discussion is expected before Release 19. However, an industry specification group on RIS at the European Telecommunications Standards Institute (ETSI) has comprehensively studied the use cases, deployment scenarios, channel models, architecture, and impact of RIS on standardization \cite{ISG_RIS_ETSI}. Interested readers can also refer to key points for potential standardisation of RISs compiled in \cite{Li-ICCC22:Considerations}. In a nutshell, with the advent of Release 18, RIS and HAPs are poised to utilize 5G-advanced/6G technology.  

Current research on HAPs has addressed their deployment strategies, operational altitude, and coverage capabilities \cite{Abbasi-MWC24:HAPS}. Meanwhile, studies on RIS have demonstrated their ability to enhance signal strength and reliability in non-terrestrial networks \cite{Ye-JPROC22:Nonterrestrial,Tekbıyık-MVT22:Reconfigurable}. The optimization of ARIS-assisted HAPs communications was presented in \cite{Gao-LWC21:Aerial}, showing ARIS can significantly improve network throughput through advanced trajectory optimization and reinforcement learning algorithms. Despite the potential benefits of TRIS, popular literature has primarily focused on ARIS. Integration of RIS with terrestrial and non-terrestrial networks is discussed in \cite{Ramezani-NET2022:Toward} with a focus on analyzing the performance of a non-orthogonal multiple access (NOMA)-based ARIS system. In \cite{Alfattani-LWC23:Resource-Efficient,Alfattani-GCWkshps22:Beyond-Cell}, a resource-efficient HAPS--RIS system that enhances beyond-cell communications was proposed, optimizing energy consumption and improving quality-of-service (QoS) through intelligent resource management. Similarly, a RIS-assisted HAPS relaying system using an amplify-and-forward protocol was presented in \cite{Odeyemi-ACCESS22:Reconfigurable}. A multi-objective optimization approach for aerodynamic HAPs equipped with RIS was introduced in \cite{Azizi-LWC23:RIS}, balancing factors like delay, trajectory, and channel variations to optimize network performance in dynamic environments.

\begin{figure}[!t]
    \centering
    \includegraphics[width=\linewidth]{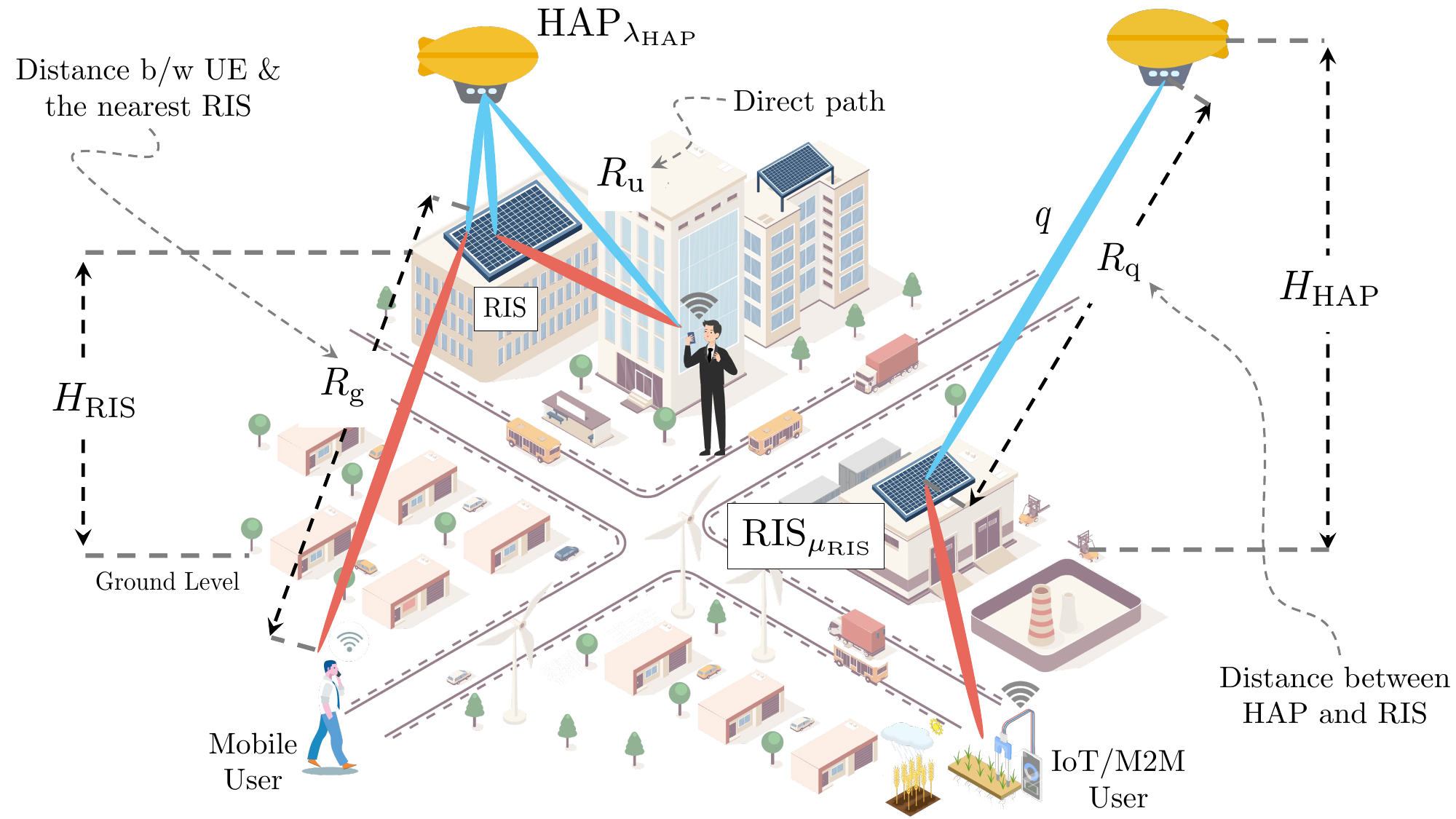}
    \caption{\textit{System model:} A network of HAPs is integrated with RISs to provide a virtual LoS communication in a dense urban environment. The HAPs and RISs are assumed to be randomly located following homogeneous PPPs. The users connect with the nearest RIS and the RISs are deployed on buildings modelled using a Boolean model \cite{blockage_urban_main}.}
    \label{fig:systemModel}
\end{figure}

This paper proposes a terrestrial RIS-enabled HAP network for densely populated areas. We employ tools from stochastic geometry to randomly model the locations of HAP and RIS as a homogeneous Poisson point process (PPP), thus capturing the essential characteristics of the system in a generalized manner. This approach allows us to evaluate the effectiveness of the proposed solution in overcoming LoS challenges and enhancing network performance in diverse environments.

The proposed system concept can be effectively deployed in several use-case scenarios. In smart cities with RISs in place as passive reflectors, HAP can periodically hover over the city to collect and offload data from IoT sensors, often deployed densely throughout the town. Similarly, in large gatherings such as sporting events, concerts, and festivals, the sudden spike in network demand within a confined geographical area can be addressed by temporarily deploying HAPs with RISs on top of buildings. This deployment provides additional backhaul capacity, alleviating congestion and ensuring seamless connectivity for attendees.

The specific contributions of this paper are threefold:
\begin{enumerate}
    \item We introduce a HAP-based network leveraging the terrestrial RIS system to improve connectivity and data offloading in smart cities and dense urban environments.
    \item Stochastic geometry is employed to model and analyze the performance of this system, considering random locations for HAPs and RISs in a 2D geometric model.
    \item We demonstrate the effectiveness of the proposed solution through theoretical analysis and simulation results, highlighting its potential to enhance network coverage and capacity significantly.
\end{enumerate}

The rest of the paper is organised as follows: Section II presents the system model and the characteristics of distance measures critical for the analysis. Section III presents the coverage and capacity performance analysis of the proposed system. Simulation results and some key observations are compiled in Section IV. Finally, the paper concludes with remarks and possible future directions, presented in Section~V.
\section{System Model}
As shown in Fig. \ref{fig:systemModel}, we consider a HAP network within urban environments characterized by dense buildings and assisted by RISs to provide enhanced connectivity through creating virtual LoS communication links. It is assumed that the HAPs are deployed in two-dimensional (2D) space at height $\HHAP$, $20\leq \HHAP \leq50$ km, following a homogeneous PPP on the $\mathbb{R}^2$ plane with density $\lambdahaps$. Similarly, the RISs are assumed to be deployed at height $\HRIS$ according to homogeneous PPP on the $\mathbb{R}^2$ plane with density $\muris$. Each RIS is equipped with $L$ reflecting elements (REs).
To evaluate the network's performance under consideration, Slivnyak's theorem~\cite{slivnyak_theor} is applied, which assumes that the user is located at the origin.
For simplicity, ideal isotropic antennas for both HAPs and users are assumed. 

Using random shape theory concepts, buildings with random locations and dimensions are modelled using a specific object process known as the Boolean model \cite{blockage_urban_main}. For the considered system, this model can be described as follows.
\begin{itemize}
    \item \textbf{Building centres:} Building centres are generated in the \(\mathbb{R}^2\) plane according to a homogeneous PPP \(\Psi_B\) with density \(\lambda_B\).
    \item \textbf{Shape parameters:} Each building is assigned unique shape parameters including length \(L_k\), width \(W_k\), and orientation \(\Theta_k\).
    \item \textbf{Probability distributions:} The lengths \(L_k\) and widths \(W_k\) of the buildings are independent and identically distributed (i.i.d.) according to specific probability density functions (PDFs) \(f_L(x)\) and \(f_W(x)\), with mean values \(\E[L]\) and \(\E[W]\), respectively. The orientation \(\Theta_k\) is a uniformly distributed random variable in \((0, 2\pi]\).
\end{itemize}

\subsection{Association Policy}
A user U can receive signals from the nearest HAP via the nearest visible RIS. The direct link between the HAP and the user is assumed to be blocked due to buildings in the urban environment. It is assumed that blockages only affect the RIS--U link and not the HAP--RIS link due to the elevated positioning of the RIS. In particular, the HAP is considered to be visible to the RIS by being positioned above a minimum height $\HMIN$, i.e., a pre-defined height above which the RISs can be served by a HAP (for simplicity, $\HRIS$ is assumed the same for all RISs in this study).

To keep the analysis comprehensive, the model also includes the case of a less expected direct path between the serving HAP and the U, for which the channel fading coefficient $u$ will be nonzero. For such cases, the role of the RIS would be to enhance the system's performance.

\subsection{Signal Model}
The received signal at the ground user can be expressed as 
\begin{align}
\label{eq:Y}
y &= A\,s+w= \bigg(\frac{\sum_{l=1}^{L} \,q_{l}\, g_{l}\,\zeta_{l}}{R_q^{\frac{\epsilon}{2}}\,\rg^{\frac{\varepsilon}{2}}} + \frac{u}{\ru^{\frac{\varrho}{2}}}\bigg)\,s+w,
\end{align}
where $s$ is the transmitted signal, $q_{l}$, $g_{l}$, and $u$ represent the fading coefficients for HAP--RIS, RIS--U, and HAP--U links, respectively, with $R_q$, $\rg$, $\ru$ denoting the corresponding distances while $\epsilon$, $\varepsilon$, $\varrho$ indicate the path-loss exponents for these links, and $w$ is the additive white Gaussian noise with zero mean and variance $N_0=\E[|w|^2]$. Furthermore, $\zeta_{l}=\exp(j\theta_{l})$ represents the response of the $l^{\mathrm{th}}$ RE in the serving RIS, with magnitude assumed to be one, and phase shift optimized to maximize the received signal-to-noise ratio (SNR) at U as $\theta_{l}=\angle{u}-\left(\angle{q_{l}}+\angle{g_{l}}\right)$. This paper assumes perfect channel state information (CSI). Assuming coherent signal combining and beamforming toward the intended user, interference from other HAPs via RISs is negligible. The user may, however, experience interference directly from the other HAPs, which can be explored in future work.

The SNR at U is given by
\begin{align}
\label{eq:snr}
  \rho&=\rho_0\,\left|\frac{\sum_{l=1}^{L} q_{l}\, g_{l}}{R_q^{\frac{\epsilon}{2}}\,\rg^{\frac{\varepsilon}{2}}}+\frac{u}{\ru^{\frac{\varrho}{2}}} \right|^2, 
\end{align}
where $\rho_0=E_s/N_0$ is the transmit SNR with $E_s=\E[|s|^2]$. 

\subsection{Distance Distributions}
To assess the system's performance, it is essential to derive the distribution of encountered horizontal distances $\omega_{q}$, $\omega_g$, and $\omega_u$  and some related statistical measures of the corresponding diagonal distances $R_q$, $\rg$, and $\ru$, respectively. 

\subsubsection{RIS--U link}
\begin{lemma} 
\label{lemma:U_RIS_distance_pdf}
The PDF of the horizontal distance $\omega_g$ from an arbitrarily located user $U$ to the nearest visible RIS is given by 
\begin{align}
\label{eq:pdf_r_gn}
f_{\omega_g}(w_g)&= 2\,\pi\,\muris\,w_g\,\exp\!\left(\!-\!\left(\Upsilon w_g+p+2\pi\muris U(w_g)\right)\right),
\end{align}
where $\Upsilon$, $p$, and $U(w_g)$ are defined respectively as
\begin{align}
&\Upsilon=\frac{2 \lambdab(\E[L]+\E[W])}{\pi},\nonumber\\
&p=\lambdab \, \E[L]\,\E[W],\nonumber\\
&U(w_g)=\frac{\exp(-p)}{\Upsilon^2}\left[1-(\Upsilon\,w_g+1)\exp(-\Upsilon\,w_g)\right].\nonumber
\end{align}

\end{lemma}\begin{IEEEproof}
The PDF in (\ref{eq:pdf_r_gn}), along with the corresponding parameters, is directly obtained from~\cite[Corollary 8.1]{blockage_urban_main}.
\end{IEEEproof}

The PDF in (\ref{eq:pdf_r_gn}) is used to find the $t^{\mathrm{th}}$ moment of the distance $\rg=\sqrt{\omega_g^2+\HRIS^2}$ in the following lemma. 

\begin{lemma}
\label{lem:mean_rg}
The $t^{\mathrm{th}}$ moment of $\rg^{-\frac{\varepsilon}{2}}$ is given by
\begin{equation}
\resizebox{1\hsize}{!}{$\begin{aligned}\label{eq:mean_rg}
&\E\Big[\rg^{\frac{-t\varepsilon}{2}}\Big]= 2\,\pi\,\muris \exp(-p)\int_{0}^{\infty}{ (w_g^2+\HRIS^2)^{-\frac{t\varepsilon}{4}}} w_g \exp\bigg(\\
&-\Big(\Upsilon w_g +2\pi\muris \frac{\exp(-p)}{\Upsilon^2}\left(1-(\Upsilon\,w_g+1)\exp(-\Upsilon\,w_g)\right)\Big)\bigg) \mathrm{d} w_g.\\
 \end{aligned}$}
\end{equation}
\end{lemma}
\begin{IEEEproof}
The expression follows from the definition of the $t^{\mathrm{th}}$ moment of a random variable. The integral in the expression can be accurately evaluated numerically.
\end{IEEEproof}

\subsubsection{HAP--RIS and HAP--U links}

\begin{proposition} 
\label{prop:pdf_rq}
The horizontal distances $\omega_q$ and $\omega_u$ have the same distribution whose PDF $(f_{\omega}(w_q)$ and $f_{\omega}(w_u))$ is given by~\cite{ppp_distribution}
\begin{align}\label{eq:pdf_rq}
f_{\omega}(w)&=2\lambdahaps\pi w\exp{(-\lambdahaps\pi w^2)},
\end{align}
for $w\geq 0$. The shorthand $\omega \in \{\omega_q,\omega_u\}$ represents both distances jointly.
\end{proposition}
\begin{IEEEproof}
Given that the user connects to the nearest HAP and the nearest visible RIS, and considering that RISs are typically densely distributed around the user, it is reasonable to assume that the serving HAP will also be nearest to the serving RIS. This is because the RIS will be very close to the user. In contrast, the distance between the user and the HAP will be much larger (tens or even hundreds of kilometres) than between the serving RIS and the user (only tens or hundreds of meters). Therefore, the horizontal distances of the HAP--U and HAP--RIS links will share the same distribution, specifically that of the nearest HAP in a homogeneous PPP. The cumulative distribution function (CDF) for this distribution is given for the plane $\mathbb{R}^2$ as $F(w)=1-\exp{(-\lambdahaps\pi w^2)}$, and its derivative yields the corresponding PDF in (\ref{eq:pdf_rq}).
\end{IEEEproof}

The PDF in (\ref{eq:pdf_rq}) is used to find the $t^{\mathrm{th}}$ moment of the propagation distances of the HAP--U and HAP--RIS links in the following lemma in which the shorthand $R \in \{R_q,\ru\}$ is used to represent both diagonal distances jointly. In particular, $R=\sqrt{\omega^2+H^2}$, with the vertical distance $H=\HHAP-\HRIS$ for the HAP--RIS link and $H=\HHAP$ for the HAP--U link.     

\begin{lemma}
\label{lem:mean_rq}
The $t^{\mathrm{th}}$ moment of $R^{-\frac{\eta}{2}}$ can be found as
\begin{align}\label{eq:mean_rqq}
&\E\Big[R^{\frac{-t\eta}{2}}\Big]=2\lambdahaps\pi  \int_{0}^{\infty}{ (w^2+H^2)^{-\frac{t\eta}{4}}}\nonumber\\
& \times w\exp{(-\lambdahaps\pi w^2)} \mathrm{d} w\nonumber\\
&=\Big(\pi\,\lambdahaps\Big) ^{\frac{\eta t}{4}} e^{\pi  H^2 \lambdahaps}  \Gamma \left(1-\frac{\eta t}{4},H^2 \lambdahaps \pi \right),
\end{align}
where $\eta \in \{\epsilon,\varrho\}$ represents the path-loss exponents of both HAP--RIS and HAP--U links, respectively, and $\Gamma(\cdot,\cdot)$ is the upper incomplete Gamma function~\cite{tableofseries}. 
\end{lemma}
\begin{figure*}[t]
\setcounter{MYtempeqncnt}{\value{equation}}
\setcounter{equation}{7}
\begin{equation}
\resizebox{.9\hsize}{!}{$\begin{aligned}
\label{eq:mean_A}
\left.\begin{matrix} \hspace{0 cm}\E[|A|]&=\overbrace{L\frac{\Gamma\left(\mu_{q}+\frac{1}{2}\right)\,\exp(-\kappa_{q}\,\mu_{q})}{\Gamma(\mu_{q})\,\left((1+\kappa_{q})\,\mu_{q}\right)^{\frac{1}{2}}}
 \frac{\Gamma\left(\mu_{g}+\frac{1}{2}\right)\,\exp(-\kappa_{g}\,\mu_{g})}{\Gamma(\mu_{g})\,\left((1+\kappa_{g})\,\mu_{g}\right)^{\frac{1}{2}}}\,_1F_1(\mu_{q}+\frac{1}{2};\mu_{q};\kappa_{q}\,\mu_{q})_1F_1(\mu_{g}+\frac{1}{2};\mu_{g};\kappa_{g}\,\mu_{g})}^{\E[\nu]}\\
 &\hspace{-2.5 cm}\times\Big(\pi\,\lambdahaps\Big) ^{\frac{\epsilon}{4}} e^{\pi  (\HHAP-\HRIS)^2 \lambdahaps}  \Gamma \left(1-\frac{\epsilon}{4},(\HHAP-\HRIS)^2 \lambdahaps \pi \right) \times \E[\rg^{\frac{-\varepsilon}{2}}]\\
 \end{matrix}\right\}\text{$P_1$} \\
 &\hspace{-16 cm}+\underbrace{ \Big(\pi\,\lambdahaps\Big) ^{\frac{\varrho}{4}} e^{\pi  \HHAP^2 \lambdahaps}  \Gamma \left(1-\frac{\varrho}{4},\HHAP^2 \lambdahaps \pi \right)  \frac{\Gamma\left(\mu_{u}+\frac{1}{2}\right)\,\exp(-\kappa_{u}\,\mu_{u})}{\Gamma(\mu_{u})\,\left((1+\kappa_{u})\,\mu_{u}\right)^{\frac{1}{2}}}
 \,_1F_1(\mu_{u}+\frac{1}{2};\mu_{u};\kappa_{u}\,\mu_{u})}_{P_2}
\end{aligned}$}
\end{equation}
\setcounter{equation}{\value{MYtempeqncnt}}
\hrulefill
\end{figure*}
\begin{figure*}[t]
\setcounter{MYtempeqncnt}{\value{equation}}
\setcounter{equation}{8}
\begin{equation}
\resizebox{.88\hsize}{!}{$\begin{aligned}
\label{eq:var_A}
&\Var[|A|]=\Bigg[\Bigg(\frac{(L^2-L)\Gamma^2\left(\mu_{q}+\frac{1}{2}\right)\,\exp(-2\,\kappa_{q}\,\mu_{q})}{\Gamma^2(\mu_{q})\,(1+\kappa_{q})\,\mu_{q}}\,\frac{\Gamma^2\left(\mu_{g}+\frac{1}{2}\right)\,\exp(-2\,\kappa_{g}\,\mu_{g})}{\Gamma^2(\mu_{g})\,(1+\kappa_{g})\,\mu_{g}}\,_1F_1^2(\mu_{q}+\frac{1}{2};\mu_{q};\kappa_{q}\,\mu_{q})\\
&\times\,_1F_1^2(\mu_{g}+\frac{1}{2};\mu_{g};\kappa_{g}\,\mu_{g})+L\Biggr) \Bigg( \Big(\pi\,\lambdahaps\Big) ^{\frac{\epsilon}{2}} e^{\pi  (\HHAP-\HRIS)^2 \lambdahaps}  \Gamma \left(1-\frac{\epsilon}{2},(\HHAP-\HRIS)^2 \lambdahaps \pi \right)\Bigg) \times \E[\rg^{-\varepsilon}] -P_1^2\Bigg]\\
& +\Bigg[ \Big(\pi\,\lambdahaps\Big) ^{\frac{\varrho}{2}} e^{\pi  \HHAP^2 \lambdahaps}  \Gamma \left(1-\frac{\varrho}{2},\HHAP^2 \lambdahaps \pi \right)-P_2^2\Bigg]. 
\end{aligned}$}
\end{equation}
\setcounter{equation}{\value{MYtempeqncnt}}
\hrulefill\\
\begin{small}
$^*$Note: $P_1$ and $P_2$ in (\ref{eq:var_A}) are defined in (\ref{eq:mean_A}).
\end{small}
\end{figure*}
\subsection{Fading Model}
Since the user is assumed to have a LoS path with the serving RIS, i.e., the nearest visible RIS, and given that both the HAP and RIS are positioned at elevated locations which guarantee a LoS path between them, Rician fading is deemed the most suitable to model both HAP--RIS and RIS--U channels. However, we use the generalized $\kappa$--$\mu$ distribution to model a more diverse set of channels that encompasses also Rician fading as a special case when $\kappa=k,\, \mu=1$ with $k$ being the Rician factor. This approach allows for a comprehensive performance analysis by accounting for a broader range of channel conditions. While Rayleigh fading is best suited to model the direct path if assumed unobstructed in urban areas, we also adopt the general $\kappa$--$\mu$ distribution to model it for broader performance analysis in which Rayleigh fading is also included as a special case when $\kappa=0,\, \mu=1$. 
Next, we approximate the distribution of end-to-end SNR in a closed form in the following Lemma. 
\begin{lemma}
\label{lem:pdf_snr}
The PDF of the end-to-end SNR can be approximated by the first term of the Laguerre series as
\begin{align}
\label{eq:pdf_snr}
    f_{\rho}(x)&\simeq\frac{1}{2\,\beta^{\alpha}\,\Gamma(\alpha)}\,\rho_0^{-\frac{\alpha}{2}}x^{\frac{\alpha-2}{2}}\exp\left({-\sqrt{\frac{x}{\beta^2\, \rho_0}}}\right).
\end{align}
Above, $\alpha=\frac{(\E[|A|])^2}{\Var[|A|]}$, $\beta=\frac{\Var[|A|]}{\E [|A|]}$, where $A$ is the combined channel response given in (\ref{eq:Y}). The mean $\E[|A|]$ and the variance $\Var[|A|]$ are provided in (\ref{eq:mean_A}) and (\ref{eq:var_A}), respectively.\stepcounter{equation} \stepcounter{equation}
\end{lemma}
\begin{IEEEproof}
See Appendix~\ref{appen:proof_pdf_snr}.
\end{IEEEproof}
%
%
\section{Performance Analysis}
In this section, the coverage probability and ergodic capacity of the RIS-based HAPs' network are derived using the statistical measures derived in the previous section.
In particular, both the coverage probability and ergodic capacity of the considered system share the same analytical form as \cite[Eq.~11]{tanash-RIS_sat} and \cite[Eq.~12]{tanash-RIS_sat}, respectively, which are rewritten in the following two theorems. However, in this context, the novel expressions for $\alpha$ and $\beta$ are used, specifically calculated using the mean and variance of the combined channel response as given in (\ref{eq:mean_A}) and (\ref{eq:var_A}), respectively, for the original system considered herein.

\begin{theorem}
 If the user connects to the nearest HAP and is assisted by the nearest visible RIS, the coverage probability can be calculated as
\begin{align}
\label{eq:cov_prob}
\operatorname{P}_c&\simeq1-\frac{\gamma\left(\alpha,\sqrt{\frac{\rho_\mathrm{th}}{\rho_0\beta^2}}\right)}{\Gamma(\alpha)},
\end{align}
where $\alpha$ and $\beta$ are defined in Lemma~\ref{lem:pdf_snr}.
\end{theorem}
The proof of (\ref{eq:cov_prob}) can be found in \cite[Theorem 1]{tanash-RIS_sat}.

\begin{theorem}
The ergodic capacity of a HAPs network assisted with RISs under the $\kappa$--$\mu$ fading can be calculated as~\cite{TCOM2}
\begin{equation}
\resizebox{1.0\hsize}{!}{$\begin{aligned}
\label{eq:capacity}
&\bar{\C}\simeq\frac{\pi}{\alpha\Gamma(\alpha)\log_e(2)}\bigg(\frac{1}{\beta^2 \rho_0}\bigg)^{{\frac{\alpha}{2}}}\csc\bigg(\frac{\pi\alpha}{2}\bigg)\, _1F_2\left(\frac{\alpha}{2};\frac{1}{2},1+\frac{\alpha}{2};-\frac{1}{4\beta^2\rho_0}\right)\\
&+\frac{\Gamma(\alpha-2)}{\Gamma(\alpha)}\frac{1}{\beta^{2}\rho_0}\, _2F_3\left(1,1;2,\frac{3}{2}-\frac{\alpha}{2},2-\frac{\alpha}{2};-\frac{1}{4\beta^2\rho_0}\right)-\frac{1}{(1 + \alpha)}\\
&\times \frac{1}{\Gamma(\alpha)}\Bigg[(2-\alpha-2\alpha^2+\alpha^3)\Gamma(\alpha-2)\Big(\log\Big(\frac{1}{\beta^2\rho_0}\Big)-2\psi^{(0)}(\alpha)\Big)\\
&+\pi\bigg(\frac{1}{\beta^2\rho_0}\bigg)^{\frac{1+\alpha}{2}}\,_1F_2\left(\frac{1}{2}+\frac{\alpha}{2};\frac{3}{2},\frac{3}{2}+\frac{\alpha}{2};-\frac{1}{4\beta^2\rho_0}\right)\sec\Big(\frac{\pi\alpha}{2}\Big)\Bigg],
 \end{aligned}$}
\end{equation}
where $\psi^{(0)}(\cdot)$ is the $0^{\mathrm{th}}$ polygamma function and $\csc(\cdot)$ is the cosecant function.
\end{theorem}
The proof of (\ref{eq:capacity}) can be found in \cite[Theorem 2]{tanash-RIS_sat}.
\begin{figure}[!t]
    \centering
    \includegraphics[width=0.99\linewidth]{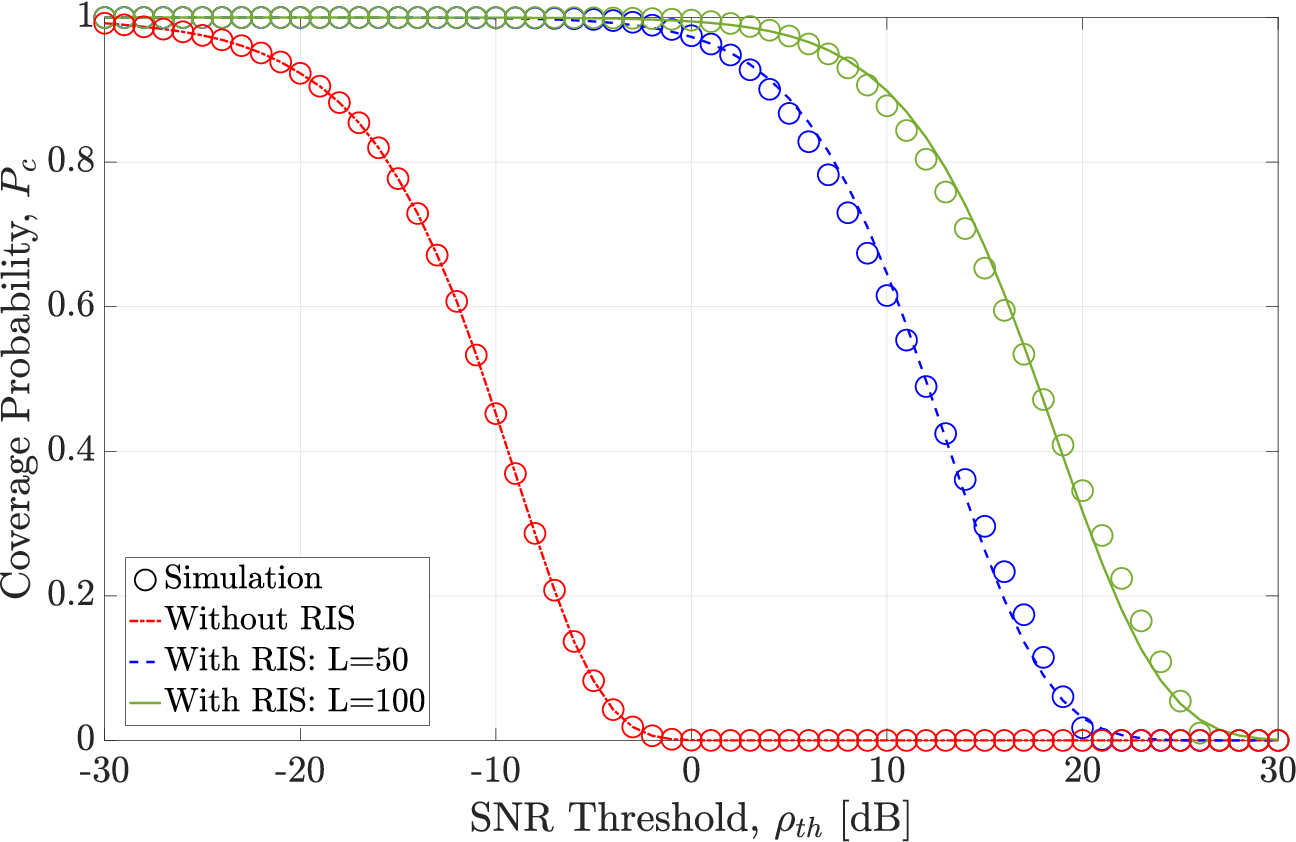}
    \caption{\textit{Effect of RIS integration with HAPS on coverage:} Coverage probability, $P_c$ vs SNR threshold, $\rho_{\mathrm{th}}$ for the cases with and without RIS links and varying number of REs.}
    \label{fig:covVsTh}
\end{figure}
\begin{figure}[!t]
    \centering
    \includegraphics[width=0.99\linewidth]{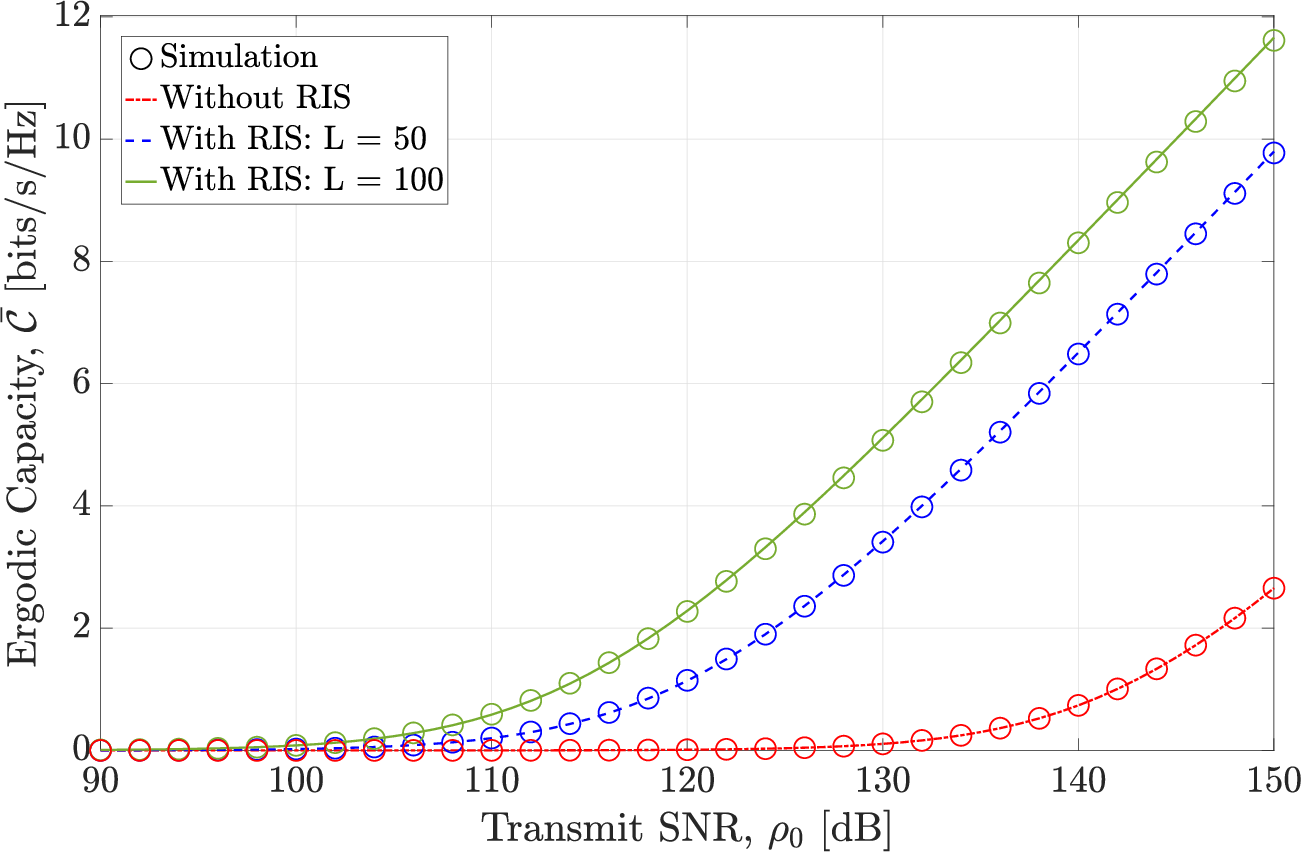}
    \caption{\textit{Effect of RIS integration with HAPS on capacity:} Ergodic capacity, $\bar{\cal{C}}$ vs transmit SNR, $\rho_0$, for the cases with and without RIS links and varying number of REs.}
    \label{fig:capVsSNR}
\end{figure}
%
\section{Results and Observations}
In this section, the performance of the proposed system is evaluated based on the derived analytical expressions and corroborated with Monte Carlo simulations. Unless stated otherwise, the parameter values have been set as follows: intensity of HAP distribution, $\lambda_{\text{HAP}} = 5 \times 10^{-6} / \mathrm{m}^{2}$; intensity of RIS distribution, $\mu_{\text{RIS}} = 50 \times 10^{-6} / \mathrm{m}^{2}$; the intensity of buildings in the urban environment, $\lambda_{\mathrm{B}} = 200 \times 10^{-6} / \mathrm{m}^{2}$; the height of HAP deployment, $H_{\mathrm{HAP}} = 50~\mathrm{km}$; height of RIS deployment, $H_{\mathrm{RIS}} = 50~\mathrm{m}$; size of the buildings, $\E[L] = \E[W] = 25~\mathrm{m}$; path loss exponents, $\epsilon = 2$, $\varepsilon = 3$, $\varrho = 3$; the parameters of fading distributions, $\kappa_u = 0$, $\mu_u = 1$, $\kappa_q = 2$, $\mu_q = 1$, $\kappa_g = 3$, $\mu_g = 1$; and $E_s = 10 \,\mathrm{W}$, $N_0=-92 \,\mathrm{dBm}$ for the link budget. The evaluation focuses on the impact of RIS integration with HAPs and its configuration on coverage probability and ergodic capacity in an urban environment.

Fig. \ref{fig:covVsTh} shows the coverage probability with respect to the increasing SNR threshold. The derived analytical results (lines in all figures) align well with the Monte Carlo simulations (circle markers in all figures), demonstrating the validity of the theoretical model. Compared to relying solely on the direct link between the HAP and the user, a significant improvement in coverage performance is observed when RIS is integrated into the system. Increasing the number of REs also further improves the system's performance. For example, at a 10 dB threshold, increasing the number of REs from $L=50$ to $L=100$ improves the coverage probability from approximately 0.6 to 0.9, marking a 50\% improvement and $6$~dB gain in SNR. Similar enhancements are observed in ergodic capacity, as depicted in Fig. \ref{fig:capVsSNR}, where integrating RIS and increasing REs lead to substantial performance gains.

\begin{figure}[!t]
    \centering
    \includegraphics[width=0.99\linewidth]{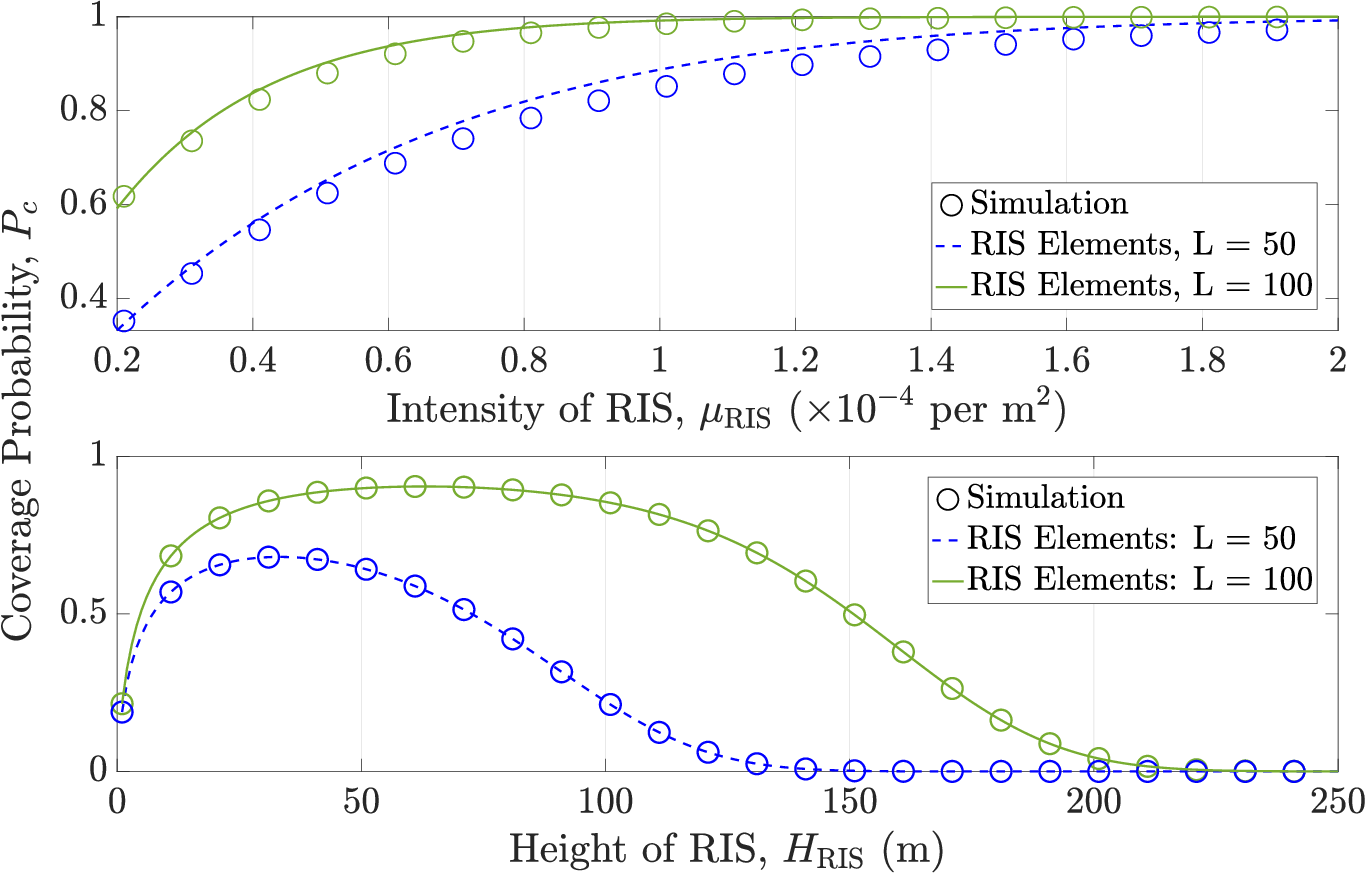}
    \caption{\textit{Effect of configuration of RIS deployment:} Coverage probability, $P_c$ vs intensity of RIS, $\mu_{\mathrm{RIS}}$, and height of RIS, $H_{\mathrm{RIS}}$ for varying number of REs.}
    \label{fig:covVsmuRIShRIS}
\end{figure}
Fig. \ref{fig:covVsmuRIShRIS} illustrates the impact of the intensity and height of RIS deployment on coverage performance. As the intensity of RIS deployment increases, the coverage probability also increases due to the higher likelihood of at least one RIS being positioned optimally to provide a LoS path to the user, thereby reducing the distance to the nearest RIS. However, beyond a certain point, further increases in RIS density result in diminishing returns as the environment becomes saturated with sufficient RIS units to ensure optimal coverage. 

The coverage probability likewise increases with the height of RIS deployment initially, reaching an optimal point before it decreases. It is because the higher RIS placement initially improves the likelihood of maintaining a LoS connection between the RIS and the user, but increased path loss due to longer distances outweighs these benefits beyond the optimal height. Thus, selecting the heights for the RISs can be traded off with the path loss to optimize the coverage and provide valuable insights for practical deployments. Coincidentally and fortunately, the observed close-to-optimal RIS, i.e., building, heights (10--100 m) are realistic and common in major cities of the world while any much higher-than-optimal heights are not so feasible as they would require a skyscraper.

\section{Conclusion}
This paper demonstrated the significant benefits of integrating RIS into HAP networks for urban environments. Stochastic geometry was employed to characterize the distribution of distances between the HAPs, RISs and users. With derived analytical expressions and Monte Carlo simulations, we have shown that RIS can effectively mitigate the impact of blockages and enhance coverage probability and ergodic capacity. Moreover, $\kappa$--$\mu$ distribution was used to model the fading, ensuring that the derived expressions can be applied to various fading scenarios by adjusting $\kappa$ and $\mu$ values. The results highlight the importance of optimizing RIS deployment height and density to achieve maximal coverage. Specifically, an optimal height for RIS deployment exists that balances path loss and line-of-sight connections, while increased RIS density improves coverage up to a saturation point. These findings provide valuable guidelines for the practical deployment of RIS in urban HAP networks and pave the way for a comprehensive analysis incorporating interference.

\appendices

\section{Proof of Lemma~\ref{lem:pdf_snr}}
\label{appen:proof_pdf_snr}
The central limit theorem can be used to approximate the combined channel response in (\ref{eq:Y}) by a normal random variable which can be further approximated by the first term of a Laguerre series (Gamma distribution) according to \cite{stochastic-book} as \begin{align}
\label{eq:pdf_A}
    f_{|A|}(x)&\simeq\frac{x^{\alpha-1}}{\beta^{\alpha}\,\Gamma(\alpha)}\,\exp\left(-\frac{x}{\beta}\right).
\end{align}
Therefore, the PDF of the end-to-end SNR defined in (\ref{eq:snr}), can be calculated using the relation $f_{\rho}(x)=\frac{1}{2 \sqrt{\rho_0 x}} f_{|A|}(\sqrt{\frac{x}{\rho_0}})$ to produce~(\ref{eq:pdf_snr}). 

The mean in (\ref{eq:mean_A}) is computed using the linearity and independence properties of expectation as $\E[{|A|}]=\E[\nu]{\E[R_q^{-\frac{\epsilon}{2}}]\E[\rg^{\frac{-\varepsilon}{2}}]}+{\E[|u|]}{\E{[\ru^{-\frac{\varrho}{2}}]}}$, where $\nu=\left|\sum_{l=1}^{L} q_{l}\, g_{l}\right|$. The expectations of the distances are obtained by substituting $t=1$ in (\ref{eq:mean_rg}) and (\ref{eq:mean_rqq}), whereas $\E[\nu]$ is provided in~\cite[Eq.~8]{tanash-RIS}, and is explicitly stated in (\ref{eq:mean_A}). The $t^{\mathrm{th}}$ moment of the direct path following $\kappa$-$\mu$ distribution is given in \cite[Eq.~3]{kappa_mu_2}. 
On the other hand, the variance in (\ref{eq:var_A}) is calculated as 
$\Var[|A|]=\Var[\nu R_q^{-\frac{\epsilon}{2}}\rg^{\frac{-\varepsilon}{2}}]+\Var[\ru^{-\frac{\varrho}{2}}|u|]$ by employing \cite[Eq.~1]{statistic}. The required first ($t=1$) and second ($t=2$) moments are calculated using (\ref{eq:mean_rg}), (\ref{eq:mean_rqq}), and \cite[Eq.~3]{kappa_mu_2}. Similarly, $\Var[\nu]$ is calculated using the defination in~\cite[Eq.~10]{tanash-RIS}.

\IEEEtriggeratref{9}

\bibliographystyle{IEEEtran}
\bibliography{IEEEabrv,ref}

\end{document}